# An XMM-*Newton*-based X-ray survey of pre-main sequence stellar emission in the L1551 star-forming complex

F. Favata[1], G. Giardino[1], G. Micela[2], S. Sciortino[2], and F. Damiani[2]

[1] Astrophysics Division – Research and Science Support Department of ESA, ESTEC, Postbus 299, NL-2200 AG Noordwijk, The Netherlands
[2] INAF – Osservatorio Astronomico di Palermo, Piazza del Parlamento 1, I-90134 Palermo, Italy

**Abstract.** We present a study of the X-ray sources present in the nearby L1551 star forming region, based on a deep XMM-*Newton* observation complemented with *Chandra* data for the brightest sources. All the known pre-main sequence stars in the region have been detected, most of them with sufficient statistics to allow a detailed study of the temporal and spectral characteristics of their X-ray emission. Significant temporal (and spectral) variability on both short and long time scales is visible for most of the stars. In particular XZ Tau shows large-amplitude variations on time scales of several hours with large changes in the intervening absorption, suggestive of the X-ray emission being eclipsed by the accretion stream (and thus of the X-ray emission being partly or totally accretion-induced). The coronal metal abundance of the WTTS sources is clustered around $Z \simeq 0.2$, while the CTTS sources span almost two orders of magnitudes in coronal $Z$, even though the photospheric abundance of all stars in the L1551 is likely to be very similar. Some individual elements (notably Ne) appear to be systematically enhanced with respect to Fe in the WTTS stars. The significant differences between the spectral and temporal characteristics of the CTTS and WTTS populations suggest that a different emission mechanism is (at least partly) responsible for the X-ray emission of the two types of stars.

## 1. Introduction

The usefulness of X-ray observations for studying star formation from the earlier stages is by now well established. Given that X-ray emission evolves strongly with age, young stars are very active X-ray sources, and X-ray surveys of young associations and clusters very effectively supplement other, more traditional approaches to membership determination in these regions (see e.g. Feigelson & Montmerle, 1999 for an extensive review of recent results in the field). For already formed, very young stars, the details of the X-ray emission mechanisms are still unclear; for the Weak-Line T Tauri (WTTS) stars, for which the disk does not any longer play a prominent role, the X-ray emitting corona is likely to be similar to the one in other, older, active stars. In the Classic T Tauri stars (CTTS) signatures of relatively massive disks are visible in the optical, UV and IR, and accretion processes often have clearly visible observational signatures. Whether accretion has a role in determining the X-ray luminosity (either enhancing or inhibiting it) is still a debated question. Some studies find no difference in X-ray properties of CTTSs and WTTSs (e.g. Feigelson et al., 2002), while other authors report significant differences between the two, with CTTSs being under-luminous in X-rays with respect to WTTSs (Flaccomio et al., 2003). Also, claims have been made that a significant fraction of the X-ray luminosity of some CTTSs is indeed due to accretion (Kastner et al., 2002), and the X-ray emission of CTTSs appears to be more time-variable with respect to WTTSs (Flaccomio et al., 2000).

The nearby L1551 star-forming cloud, at a distance $d \simeq 140$ pc (Kenyon et al., 1994), is a well studied site of star formation, and its population has been extensively characterized. Low-mass recently formed stars have been searched in the cloud through their H$\alpha$ emission (Jones & Herbig, 1979; Feigelson & Kriss, 1983; Briceño et al., 1993), through their proper motion (Jones & Herbig, 1979; Gomez et al., 1992), and through their X-ray emission (Feigelson & Decampli, 1981; Feigelson et al., 1987; Carkner et al., 1996). Most of the phenomenology of low-mass star formation is present in the cloud, including highly embedded sources (such as IRS5), CTTSs, WTTSs, and proto-stellar outflows (Herbig Haro objects).

In this paper we present a deep (nominal 50 ks) XMM-*Newton* observation of the L1551 star-forming complex. For the X-ray bright sources we have also analyzed the *Chandra* ACIS observation of the same region obtained from the public archive, limiting ourselves to the temporal and spectral analysis of these few sources. An initial study of all the *Chandra* sources in the region is presented by Bally et al. (2003). The XMM-*Newton* observation presented here has been also discussed by Favata et al. (2002), who did focus onto the X-ray emission from the Herbig-Haro object HH 154.

The present paper is organized as follow. In Sect. 2 the XMM-*Newton* and *Chandra* observations are presented, with their analysis; in Sect. 3 we briefly discuss the results for the X-ray sources which are not known to be pre-main sequence stars, while in Sect. 4 the results for each known pre-main sequence star are presented. These results are discussed in Sect. 5.

## 2. Observations

### 2.1. XMM-Newton

The XMM-*Newton* observation discussed in this paper consists in a deep (50 ks nominal) exposure of the L1551 star-forming cloud, obtained starting on Sept. 9 2000 at 19:10 UTC. All three EPIC cameras were active at the time of the observation, in full-frame mode with the medium filters.

Data have been processed by us with the standard SAS V5.3.1 pipeline system, concentrating, for the spectral and timing analysis, on the EPIC-pn camera. As most of the background in XMM-*Newton* observations is concentrated in individual, relatively short-lasting proton flares, we have retained only time intervals in which the count rate for the whole frame of photons above 8 keV was below a certain threshold (0.5 cts/s in the present case). This effectively reduces the background by a factor of $\simeq 4$ while omitting only $\simeq 5\%$ of the observing time. The effective duration of the observation was $\simeq 56.8$ ks for the MOS cameras and $\simeq 54.5$ ks for the pn camera, which were reduced to $\simeq 55$ ks and 51 ks respectively after filtering out the high background intervals. In order to minimize the unwanted contribution of non-X-ray events we have retained for analysis only the X-ray photons whose energy is in the 0.3–7.9 keV range. The details of this process are described in a previous paper (Favata et al., 2002) and will not be repeated here.

Source detection (again as described in detail in Favata et al., 2002) was performed on the summed data set obtained with the two MOS and one pn EPIC cameras, using the Wavelet Transform detection algorithm developed at Palermo Astronomical Observatory (PWDETECT, Damiani et al. 2002 in preparation). The L1551 observation has been taken with the medium filters; in such a case we have derived that the value of the relative efficiency of the pn and of the individual MOS cameras is 2.94, hence the summed data set has a single MOS-equivalent cleaned exposure time of $2 \times 55 + 2.94 \times 51 = 260$ ks. The threshold for source detection has been taken as to ensure a maximum of one spurious source per field.

For the 27 sources with sufficient $S/N$ (more than 50 counts) spectral and timing analysis was performed on the EPIC pn data. Source and background photons were extracted using a set of scripts purposely developed at Palermo Observatory. Source and background regions were defined interactively in the DS9 display software, with the background extracted from regions on the same CCD chip and at the same off-axis angle as for the source region. Response matrices ("RMF and ARF files") appropriate for the position and size of the source extraction regions were computed. The spectral analysis has been performed using the XSPEC package, after rebinning the source spectra to a minimum of 20 source counts per (variable width) spectral bin.

#### 2.1.1. Combined pn+MOS1+MOS2 data

A total of 81 X-ray sources were detected using PWDETECT. For each source we report the source position, the count rate for the combined cameras pn+MOS1+MOS2, and the presence of a possible counterpart in the USNO catalog and in SIMBAD[1]. The merged EPIC pn+MOS1+MOS2 image is shown in Fig. 1. In Table 1 also the distance between the X-ray source position and the position of the corresponding (if any) object in the USNO catalog or in SIMBAD is given, together with the $R$ magnitude $B - R$ color from the USNO catalog. The "Id" column provides the name of the potential counterpart found in the SIMBAD catalog by searching within a radius of 15 arcsec from the X-ray source position.

Among the 81 sources detected within the combined data set (pn, MOS1, MOS2), 27 have enough counts in the pn camera for their spectra to be analyzed. The X-ray image is shown in Fig. 1. The characteristics of the X-ray bright sources are listed in Table 2, the last column in Table 1 giving the number of the corresponding source in Table 2.

All X-ray sources with a known counterpart have been detected with enough photons to allow for their pn spectrum to be studied, with the exception of sources 12, 28, 73, 77 and 80. Sources 77 and 80 fall outside the field of view of the pn camera (and their faintness and large offset angle prevent their MOS spectrum to be usefully analyzed), while sources 12 and 73 are too weak for a reliable spectrum to be extracted from the pn camera. In addition, source 28 lies in a region of high background due to its proximity to source 29 (the bright T Tau star HL Tau). Of these 5 sources, source 28 (CoKu Tau 2, or LkH$\alpha$ 358) is a known late-type pre-main sequence star (Bertout, 1989), while source 77 ([CFK96] RX29) has been detected in X-rays by Carkner et al. (1996), who report that the star has no H$\alpha$ emission or Li I $\lambda$6707 absorption (although the $S/N$ of their spectra is low). However, it falls among the T Tau stars in the X-ray/optical brightness diagram and has proper motion not inconsistent with the cloud, so that they classify it as possible newly detected T Tauri in the star-forming complex.

#### 2.1.2. The EPIC-pn data

Table 2 gives the background-corrected count rate in the 0.3–7.9 keV band for the pn camera and, for the sources previously detected with ROSAT, the table provides ROSAT PSPC count rates as derived from the study by Carkner et al. (1996). The name of the optical counterpart is also given.

---

[1] This field is still not available in the 2MASS catalog so that no IR data are publicly available.

**Table 1.** XMM-*Newton* sources in the L1551 cloud region, as detected in the combined EPIC pn+MOS1+MOS2 image. If a counterpart is present in the SIMBAD catalog its name is listed in the "Id" field. If only a magnitude is given, the counterpart is only present in the USNO A2 catalog. Counterparts have been searched within a radius of 15 arcsec from the X-ray source position. $R$ magnitudes and $B-R$ colors are from the USNO catalog. The X-ray count rate is expressed in "MOS equivalent" units. $r$ is the distance between the X-ray source and the candidate counterpart. The last column gives the number of the spectrum of the source as used in Table 2.

| Source | R.A. (J2000) | Dec. (J2000) | Count rate $(\text{ks})^{-1}$ | $r$ arcsec | $R$ | $B-R$ | Id | Spec. |
|---|---|---|---|---|---|---|---|---|
| 1 | 4 30 53.60 | +18 00 01.1 | 1.95±0.35 | – | – | – | – | – |
| 2 | 4 31 01.40 | +18 01 53.1 | 1.68±0.28 | – | – | – | – | – |
| 3 | 4 31 02.75 | +18 10 48.0 | 0.38±0.10 | – | – | – | – | – |
| 4 | 4 31 03.71 | +17 57 44.7 | 1.87±0.34 | – | – | – | – | – |
| 5 | 4 31 04.19 | +18 07 23.3 | 3.55±0.30 | – | – | – | – | S27 |
| 6 | 4 31 05.66 | +18 03 22.6 | 33.23±0.92 | 2.1 | – | – | JH188 | S26 |
| 7 | 4 31 08.36 | +18 01 36.0 | 0.98±0.19 | – | – | – | – | – |
| 8 | 4 31 11.09 | +18 21 58.7 | 5.40±1.68 | – | – | – | – | – |
| 9 | 4 31 11.37 | +18 17 21.2 | 0.81±0.14 | – | – | – | – | – |
| 10 | 4 31 15.63 | +18 21 07.8 | 0.37±0.11 | – | – | – | – | – |
| 11 | 4 31 15.91 | +18 20 09.0 | 10.27±0.54 | 2.5 | – | – | [BHS98] MHO 9 | S25 |
| 12 | 4 31 16.94 | +17 58 02.8 | 7.83±1.49 | 9.5 | – | – | [GRL2000] 10 | – |
| 13 | 4 31 17.21 | +18 02 19.8 | 0.81±0.17 | – | – | – | – | – |
| 14 | 4 31 17.31 | +18 09 28.2 | 0.54±0.10 | – | – | – | – | – |
| 15 | 4 31 21.21 | +17 59 49.8 | 0.89±0.21 | – | – | – | – | – |
| 16 | 4 31 23.32 | +18 08 07.1 | 0.81±0.11 | – | – | – | – | S24 |
| 17 | 4 31 24.09 | +18 00 24.9 | 8.36±0.48 | 3.2 | – | – | [BHS98] MHO 4 | S23 |
| 18 | 4 31 24.58 | +18 11 40.9 | 0.93±0.12 | – | – | – | – | – |
| 19 | 4 31 25.32 | +18 16 19.3 | 618.22±3.20 | 2.4 | – | – | HD 285845 | S22 |
| 20 | 4 31 26.88 | +18 07 58.8 | 2.37±0.18 | – | – | – | – | S21 |
| 21 | 4 31 29.03 | +18 03 38.7 | 0.84±0.18 | – | – | – | – | – |
| 22 | 4 31 30.18 | +18 20 42.1 | 1.73±0.25 | – | – | – | – | – |
| 23 | 4 31 30.21 | +18 05 33.4 | 0.34±0.08 | – | – | – | – | – |
| 24 | 4 31 30.24 | +17 56 31.6 | 0.50±0.14 | – | – | – | – | – |
| 25 | 4 31 30.78 | +17 59 06.6 | 1.06±0.19 | – | – | – | – | – |
| 26 | 4 31 34.04 | +18 08 07.2 | 1.40±0.15 | 6.3 | – | – | HH 154 | S20 |
| 27 | 4 31 34.04 | +18 02 19.3 | 1.05±0.30 | – | – | – | – | – |
| 28 | 4 31 36.06 | +18 13 43.9 | 0.84±0.12 | 0.7 | 16.9 | 2.2 | CoKu Tau 2 | – |
| 29 | 4 31 38.32 | +18 13 59.4 | 19.06±0.51 | 2.3 | – | – | HL Tau | S19 |
| 30 | 4 31 38.63 | +18 16 17.0 | 0.45±0.09 | – | – | – | – | – |
| 31 | 4 31 39.99 | +18 13 59.2 | 135.09±1.35 | 1.4 | – | – | XZ Tau | S18 |
| 32 | 4 31 40.34 | +18 12 12.5 | 6.06±0.27 | – | – | – | – | – |
| 33 | 4 31 40.82 | +18 17 35.3 | 0.44±0.10 | – | – | – | – | – |
| 34 | 4 31 43.31 | +18 03 18.9 | 0.76±0.15 | – | – | – | – | – |
| 35 | 4 31 44.01 | +18 10 34.7 | 1.16±0.12 | 1.5 | – | – | [SB86] L1551 2 | S16 |
| 36 | 4 31 44.87 | +18 10 06.7 | 0.35±0.08 | – | – | – | – | – |
| 37 | 4 31 49.74 | +18 13 05.2 | 2.89±0.25 | – | – | – | – | S15 |
| 38 | 4 31 51.43 | +18 14 58.7 | 0.69±0.15 | – | – | – | – | – |
| 39 | 4 31 54.13 | +18 08 24.6 | 0.48±0.11 | 11.4 | 18.6 | 0.6 | – | – |
| 40 | 4 31 55.24 | +18 01 20.7 | 0.91±0.17 | – | – | – | – | – |
| 41 | 4 31 55.75 | +18 07 07.7 | 0.35±0.08 | – | – | – | – | – |
| 42 | 4 31 56.05 | +17 59 09.0 | 1.45±0.29 | 12.9 | 19.3 | 0.8 | – | – |
| 43 | 4 31 57.54 | +18 09 29.8 | 0.24±0.07 | 13.6 | 17.9 | 1.9 | – | – |
| 44 | 4 31 58.02 | +18 21 38.0 | 80.44±1.58 | 3.7 | – | – | V710 Tau A,B | S14 |
| 45 | 4 31 58.72 | +18 18 42.5 | 1.55±0.22 | 1.3 | – | – | [FK83] LDN1551 | S13 |
| 46 | 4 31 59.26 | +18 16 59.6 | 1.89±0.20 | 5.1 | – | – | [SB86] L1551 3 | S12 |
| 47 | 4 31 59.48 | +18 13 08.9 | 1.18±0.14 | – | – | – | – | S11 |
| 48 | 4 31 59.53 | +18 01 12.1 | 1.37±0.20 | – | – | – | – | – |
| 49 | 4 32 01.30 | +18 05 02.2 | 6.25±0.34 | – | – | – | – | S10 |
| 50 | 4 32 01.30 | +18 06 59.6 | 0.48±0.10 | – | – | – | – | – |
| 51 | 4 32 02.17 | +18 09 11.9 | 0.15±0.05 | – | – | – | – | – |
| 52 | 4 32 02.73 | +17 58 38.6 | 2.77±0.35 | – | – | – | – | – |

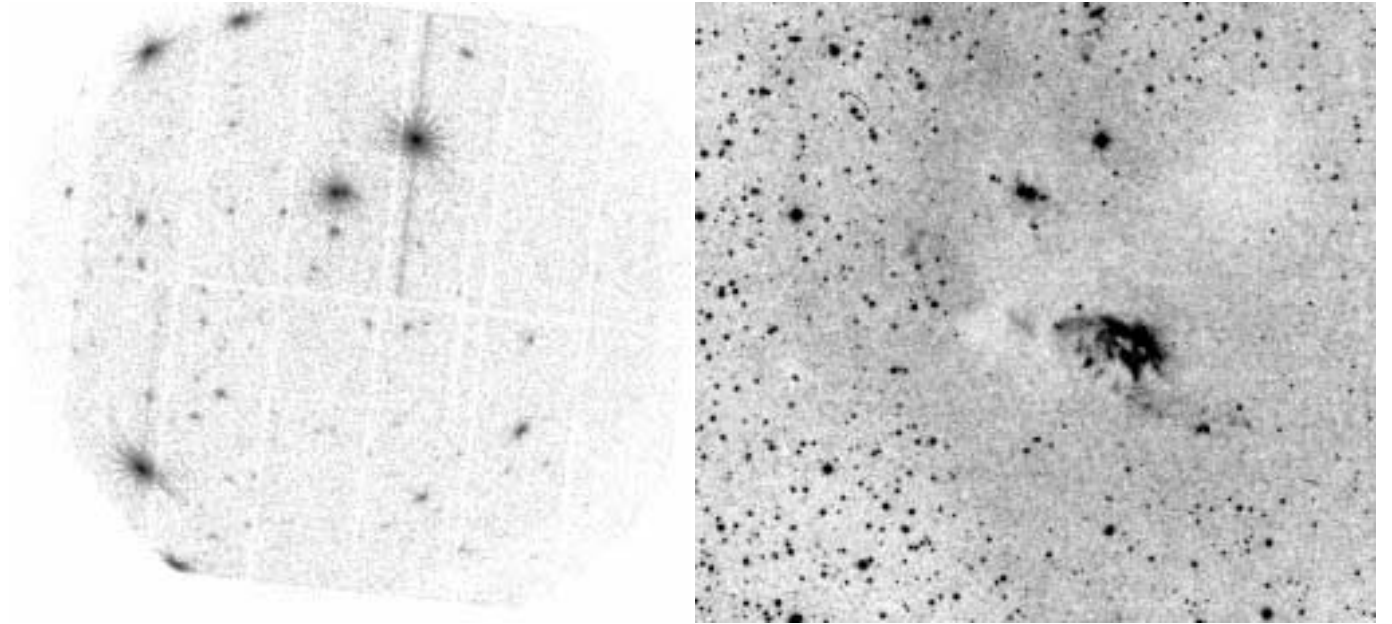

**Fig. 1.** Left panel: merged EPIC pn+MOS1+MOS2 image of the L1551 region. Right panel: the same region (to scale) in the red Palomar plates (from the Digital Sky Survey). The bright nebulosity near the center of the optical image is the outflow from the IRS5 embedded protostar.

**Table 1.** *(continued)* XMM-*Newton* sources in the L1551 cloud region.

| Source | R.A. (J2000) | Dec. (J2000) | Count rate $(\text{ks})^{-1}$ | $r$ arcsec | $R$ | $B-R$ | Id | Spec. |
|---|---|---|---|---|---|---|---|---|
| 53 | 4 32 04.50 | +18 09 38.73 | 0.92±0.13 | – | – | – | – | – |
| 54 | 4 32 04.51 | +18 08 14.39 | 1.89±0.19 | – | – | – | – | S9 |
| 55 | 4 32 05.05 | +18 00 00.68 | 0.52±0.14 | 1.4 | 12.9 | 1.6 | – | – |
| 56 | 4 32 05.63 | +18 05 41.53 | 1.00±0.16 | – | – | – | – | – |
| 57 | 4 32 06.03 | +18 03 59.87 | 4.37±0.32 | – | – | – | – | S8 |
| 58 | 4 32 07.48 | +18 00 06.91 | 0.43±0.13 | – | – | – | – | – |
| 59 | 4 32 09.40 | +17 57 24.62 | 147.28±2.66 | 1.8 | – | – | V1075 Tau | S7 |
| 60 | 4 32 10.58 | +18 00 36.43 | 1.24±0.28 | – | – | – | – | – |
| 61 | 4 32 10.59 | +18 17 2.69 | 0.25±0.08 | 12.9 | 18.6 | 0.6 | – | – |
| 62 | 4 32 10.89 | +18 16 23.99 | 0.52±0.12 | – | – | – | – | – |
| 63 | 4 32 11.53 | +18 12 42.12 | 1.25±0.18 | 9.8 | 18.1 | 0.8 | – | – |
| 64 | 4 32 12.73 | +18 08 08.31 | 0.65±0.15 | – | – | – | – | – |
| 65 | 4 32 13.72 | +18 14 06.65 | 0.32±0.09 | – | – | – | – | – |
| 66 | 4 32 14.16 | +18 20 08.55 | 417.76±7.15 | 8.9 | – | – | V827 Tau | S6 |
| 67 | 4 32 14.38 | +18 04 55.38 | 3.28±0.30 | – | – | – | – | S5 |
| 68 | 4 32 15.88 | +18 01 39.36 | 478.82±3.85 | 1.0 | – | – | V826 Tau | S3 |
| 69 | 4 32 15.92 | +18 10 44.51 | 4.70±0.33 | – | – | – | – | S4 |
| 70 | 4 32 16.06 | +18 12 48.64 | 12.52±0.54 | 1.4 | – | – | [BHS98] MHO 5 | S2 |
| 71 | 4 32 16.34 | +18 10 57.55 | 0.37±0.11 | – | – | – | – | – |
| 72 | 4 32 19.88 | +18 11 00.39 | 1.69±0.23 | 2.0 | 14.5 | 1.7 | – | S1 |
| 73 | 4 32 21.52 | +18 12 58.58 | 1.14±0.20 | 1.8 | – | – | HD 285844 | – |
| 74 | 4 32 25.00 | +18 07 13.91 | 0.85±0.17 | – | – | – | – | – |
| 75 | 4 32 25.62 | +18 10 26.11 | 1.58±0.42 | – | – | – | – | – |
| 76 | 4 32 26.22 | +18 11 44.23 | 0.59±0.17 | – | – | – | – | – |
| 77 | 4 32 26.83 | +18 18 26.85 | 6.10±1.65 | 3.5 | – | – | [CFK96] RX29 | – |
| 78 | 4 32 28.01 | +18 09 02.23 | 2.03±0.44 | – | – | – | – | – |
| 79 | 4 32 28.10 | +18 06 53.82 | 0.86±0.18 | – | – | – | – | – |
| 80 | 4 32 29.45 | +18 13 59.82 | 20.61±1.85 | 0.4 | – | – | RX J0432.4+1814 | – |
| 81 | 4 32 37.27 | +18 09 47.67 | 3.08±0.92 | 13.6 | 17.2 | 1.4 | – | – |

**Table 2.** X-ray bright XMM-*Newton* sources in L1551 for which spectral and timing analysis has been performed from EPIC pn data. The ROSAT count rate is as given in Carkner et al. (1996). In the ROSAT PSPC observation HL Tau is unresolved from the nearby XZ Tau, as discussed in the text.

| Source | EPIC-pn rate $(\mathrm{ks})^{-1}$ | ROSAT rate $(\mathrm{ks})^{-1}$ | Counterpart |
|---|---|---|---|
| S1 | $1.1 \pm 0.2$ | – | USNO-A2.0 star |
| S2 | $11.1 \pm 0.6$ | – | [BHS98] MHO 5 |
| S3 | $267.0 \pm 2.5$ | 103 | V826 Tau |
| S4 | $4.1 \pm 0.4$ | – | – |
| S5 | $3.9 \pm 0.5$ | – | – |
| S6 | $36.7 \pm 1.0$ | 67 | V827 Tau |
| S7 | $69.2 \pm 1.3$ | 47 | V1075 Tau |
| S8 | $2.5 \pm 0.5$ | – | – |
| S9 | $2.3 \pm 0.4$ | – | – |
| S10 | $3.6 \pm 0.5$ | – | – |
| S11 | $3.2 \pm 0.4$ | – | – |
| S12 | $3.0 \pm 0.4$ | – | [SB86] L1551 3 |
| S13 | $1.5 \pm 0.3$ | – | [FK83] LDN 1551 9 |
| S14 | $45.6 \pm 1.1$ | 54 | V710 Tau A,B |
| S15 | $1.6 \pm 0.2$ | – | – |
| S16 | $2.0 \pm 0.5$ | – | [SB86] L1551 2 |
| S17 | $7.2 \pm 0.6$ | – | – |
| S18 | $132.8 \pm 1.7$ | 20 | XZ Tau |
| S19 | $40.1 \pm 1.0$ | unres. | HL Tau |
| S20 | $1.3 \pm 0.3$ | – | HH 154 |
| S21 | $3.0 \pm 0.4$ | – | – |
| S22 | $538.9 \pm 3.5$ | 151 | HD 285845 |
| S23 | $5.2 \pm 0.6$ | 3.1 | [BHS98] MHO 4 |
| S24 | $1.7 \pm 0.3$ | – | – |
| S25 | $5.3 \pm 0.6$ | – | [BHS98] MHO 9 |
| S26 | $22.3 \pm 0.8$ | 9.3 | JH 188 |
| S27 | $2.2 \pm 0.3$ | – | – |

## 2.2. Chandra *data*

The *Chandra* ACIS observation of the L1551 cloud was taken starting on July 23 2001, with a nominal exposure time of 80 ks. The data were retrieved from the public data archive, with no re-processing done on the archival data. Source and background regions for the four bright X-ray sources discussed here (XZ Tau, HL Tau, HD 285845 and JH 188) were defined in DS9, and light curves and spectra were extracted from the cleaned photon list using CIAO V. 2.2.1 threads, which were also used for the generation of the relative response matrices. Spectral analysis was performed in XSPEC in the same way as for the XMM-*Newton* spectra.

The *Chandra* observation is discussed in detail by Bally et al. (2003), who list all the sources detected within the *Chandra* field. All bright X-ray sources listed in Table 2 are also present in the list of *Chandra* sources, except for the ones which fall outside the *Chandra* field of view.

The results of the spectral analysis of the *Chandra* spectra performed here consistently result in very high absorbing column density, significantly higher than the one derived from the XMM-*Newton* EPIC-pn data (which, for the foreground sources, is in fair agreement with the interstellar column density expected on the basis of the source's distance). This is consistent with the known presence of (likely carbon-based) contamination on the ACIS chips, causing additional low-energy absorption (up to 50% near the C edge) not accounted for in the current response matrices (Plucinsky et al., 2002). Therefore the $N(H)$ values derived from the *Chandra* spectra are known to be consistently overestimated and will not be discussed further in this work.

## 3. Spectral and timing analysis of sources unrelated to the L1551 cloud

Of the 27 X-ray brighter sources detected in the XMM-*Newton* observation, 17 have a cataloged optical or radio counterpart. Of these, 9 are previously known pre-main sequence (PMS) stars and will be discussed in detail in the Sect 4, while the remaining 8 without a previously known PMS counterpart are discussed here together with the bright X-ray sources without any visible counterpart. The best-fit parameters for the spectral analysis of all the sources are reported in Table 3.

**Source S1** This faint X-ray source has a clearly visible optical counterpart in the Palomar plates (which appears in the USNO-A2.0 catalog with number 1050-01299961) at a distance of 1.8 arcsec from the X-ray source centroid. Its X-ray spectrum cannot be satisfactorily fit with a power-law, while it is well described ($P = 0.45$, where $P$ is the null hypothesis probability of the fit) by an absorbed thermal spectrum with $N(\mathrm{H}) = (0.89 \pm 0.65) \times 10^{22}$ cm$^{-2}$ and $kT \leq 1.0$ keV. Using the USNO $B$ and $R$ magnitudes (respectively 16.2 and 14.5) and the best fit value of $N(\mathrm{H})$ the intrinsic $B - R$ index of the source was derived. For $N(\mathrm{H}) \gtrsim 2 \times 10^{21}$, the intrinsic $B - R$ implies a main sequence star hotter than a F8, shining through the L1551 cloud, at a distance greater than $\sim 900$ pc.

**Source S10** This faint X-ray source does not have a cataloged counterpart in SIMBAD or the USNO A2.0 catalog. Nevertheless it appears to have a very faint counterpart in the digitized red Palomar plate, which Bally et al. (2003) identify as a star. The XMM-*Newton* spectrum is very hard and cannot be fit with an absorbed thermal emission model, but, as shown in Table 3, it is well described by an absorbed power law with a spectral index $\gamma = 1.1 \pm 0.5$, pointing to the likelihood of the source being of extra-galactic origin (AGN or the like).

**Source S12** This faint X-ray source lies at 5 arcsec from [SB86] L1551 3, a radio source detected by Snell & Bally (1986) in their study of compact radio sources associated with molecular outflows. [SB86] L1551 3 has no IR counterpart and no radio spectral information is available, so that the authors exclude it from the list of likely radio-emitting stellar sources embedded within the molecular cloud. Giovanardi et al. (2000) have also made radio observations of this field; beside L1551 IRS 5, they detect 25 other radio sources including [SB86] L1551 3. From extra-galactic source counts, they conclude that most of these radio sources are likely to be of extra-galactic origin. The X-ray spectrum can in principle be fit with a thermal spectrum, with a very high best fit temperature ($kT = 44$ keV), but it is also well described by an absorbed power-law spectrum with spectral index $\gamma = 1.2 \pm 0.4$ (see Table 3). This, together with the absence of an optical counterpart in the Palomar plate indicates that this source is most likely extra-galactic.

Source S13 This faint X-ray source lies within 1.3 arcsec from the GSC star 01269-00641, which corresponds to star [FK83] LDN 1551 9 in Feigelson & Kriss (1983). They classify it as K6 and exclude it from their list of potential pre-main sequence stars since it does not show H$\alpha$ emission (EW(H$\alpha$)= $-0.6 \pm 0.5$). This however would not rule it out as a WTTS, which may not show H$\alpha$ emission. By using the values provided by Feigelson & Kriss (1983) for the magnitude, $V = 13.5$, and absorption, $A_V = 0.7$, we derive for this star a photometric parallax of $\sim 120$ pc, which is not inconsistent with the star being part of the L1551 star-forming complex. The X-ray spectrum of the star is well described ($P = 0.89$) by an absorbed one temperature spectrum with $N(\text{H}) = 0.15 \pm 0.19 \times 10^{22}$ cm$^{-2}$ and $kT = 1.05 \pm 0.17$ (with a fixed relative metal abundance at $Z = 0.3$)[2]. The flux of the star in the 0.35–7.5 keV band is $7.09 \times 10^{-15}$ erg cm$^{-2}$ s$^{-1}$. Since the absorbing column density is low, this value can be used directly with $V$ to derive a ratio of X-ray flux over optical flux of $f_X/f_V \sim 5 \times 10^{-4}$, typical of a relatively high-activity coronal source, not inconsistent with its being a pre-main sequence star. Further optical observations are needed in order to establish the nature of this star and its location with respect to the cloud.

Source S16 This faint X-ray source lies at 1.5 arcsec from [SB86] L1551 2b, a component of the radio source [SB86] L1551 2 detected by Snell & Bally (1986). [SB86] L1551 2 has a non-thermal radio spectrum ($\gamma = -1.4$) and no IR counterpart therefore the authors classifies it as probable extra-galactic object. The source does not have an optical counterpart in the Palomar plates, and its X-ray spectrum is well described by an absorbed power law model with a spectral index $\gamma = 1.6 \pm 0.7$, while it cannot be satisfactorily fit by a thermal model, so that the source is most likely extra-galactic.

Source S20 This is the X-ray source associated with HH 154, the proto-stellar jet emanating from the L1551 IRS5 protostar, discussed in detail by Favata et al. (2002).

Source S22 This bright X-ray source is identified with the the binary system HD 285845 which lies at 2.4 arcsec from the X-ray source. The star, also known as V1073 Tau, is excluded from cloud membership by its radial velocity and proper motion by Walter et al. (1988). According to the same study the primary star of the system does not appear to be a PMS since it does not show detectable Li I $\lambda$6707. The spectral type is G8, with color indices $U - B = 0.23$ and $B - V = 0.77$, consistent with the ones of a main sequence star, and a projected rotational velocity of $v \sin i = 30$ km/s (Walter et al., 1988). Schneider et al. (1998) report, using HST FGS observations, a separation for the binary companion of 73 mas and a magnitude difference among the components of 1.19 mag. The apparent magnitude $V = 10.28$ implies a photometric parallax of 90 pc, indicating that HD 285845 likely is an active binary system in the foreground of L1551. Its mean intrinsic X-ray luminosity in the XMM-*Newton* observation (at the assumed distance of 90 pc) is $1.86 \times 10^{30}$ erg s$^{-1}$, typical of active binary stars.

HD 285845 shows remarkable X-ray variability during the XMM-*Newton* observation, with its flux decreasing by a factor $\sim 2$ in less than 30 ks (Fig. 2), in what could be interpreted as the decay of a long-lasting flare. However, the light curve is far from being a simple exponential decay, with very significant shorter-term variability superimposed over a longer-term decaying trend. The good statistics of the source allows the variability to be studied in detail.

The X-ray spectrum cannot be satisfactorily fit with a "classic" 2 temperature plasma model ($P \sim 10^{-7}$). The presence of large fit residuals at energies where metal lines are expected suggests that a 2 temperature plasma model with varying metal abundances may provide a better description of the spectrum of HD 285845. This is indeed the case; as shown in Fig. 2 a model with individually varying metal abundances provide a good fit to the source spectrum ($P = 0.31$), with comparable emission measure for the two temperature components. The best-fit values[3] of the model parameters are summarized in Table 8 and indicate a significant over-abundance of Ne and Ca. The best-fit value for $N(\text{H})$ is low and therefore consistent with the source being in front of the L1551 cloud.

As shown in Fig. 2, the light curve of HD 285845 during the *Chandra* observation also shows significant variability on typical time scales comparable to the ones apparent in the XMM-*Newton* data. The *Chandra* spectrum also requires individually varying metal abundances to be fit, with best-fit metal abundances very similar to the ones derived from the XMM-*Newton* data, with the only exception of Ca, for which the ACIS data imply an upper limit of 1, in contrast with the high value (as evident by the strong Ca line visible in the spectrum) derived from the XMM-*Newton* spectrum. An other discrepancy present between the XMM-*Newton* and *Chandra* spectra is the absorbing column density, in agreement with the consistently too high $N(H)$ values derived on ACIS spectra (see Sect. 2.2).

Source S26 This X-ray source lies at 2.1 arcsec from the star JH 188, which appears in the Jones & Herbig (1979) proper motion catalog of T Tau variables and other stars associated with the Taurus-Auriga dark clouds. The star is characterized by a large proper motion, significantly different from the cloud members' typical motion. No other references to JH 188 are present in SIMBAD. Nevertheless this bright X-ray source ap-pears to be source 13 observed with ROSAT by Carkner et al. (1996), which they identify with LP 415-1165, a M2 star in the Luyten proper motion catalog of 1971, which presumably is a foreground dwarf unrelated to the cloud again because of its large motion (Cudworth & Herbig, 1979). We conclude that JH 188 and LP 415-1165 are most likely the same star.

---

[2] Our best-fit value of $N(\text{H})$ is consistent with the value for $A_V$ given by Feigelson & Kriss (1983), which corresponds to $N(\text{H}) = 0.13 \pm 0.06 \times 10^{22}$ cm$^{-2}$, using a conversion factor $N(\text{H})/A_V = 1.9 \times 10^{21}$ atoms cm$^{-2}$ mag$^{-1}$ (see e.g. Cox, 2000).

[3] See Sect. 4 for the details on the procedure followed in order to fit the spectral data with a thermal model with varying individual metal abundances.

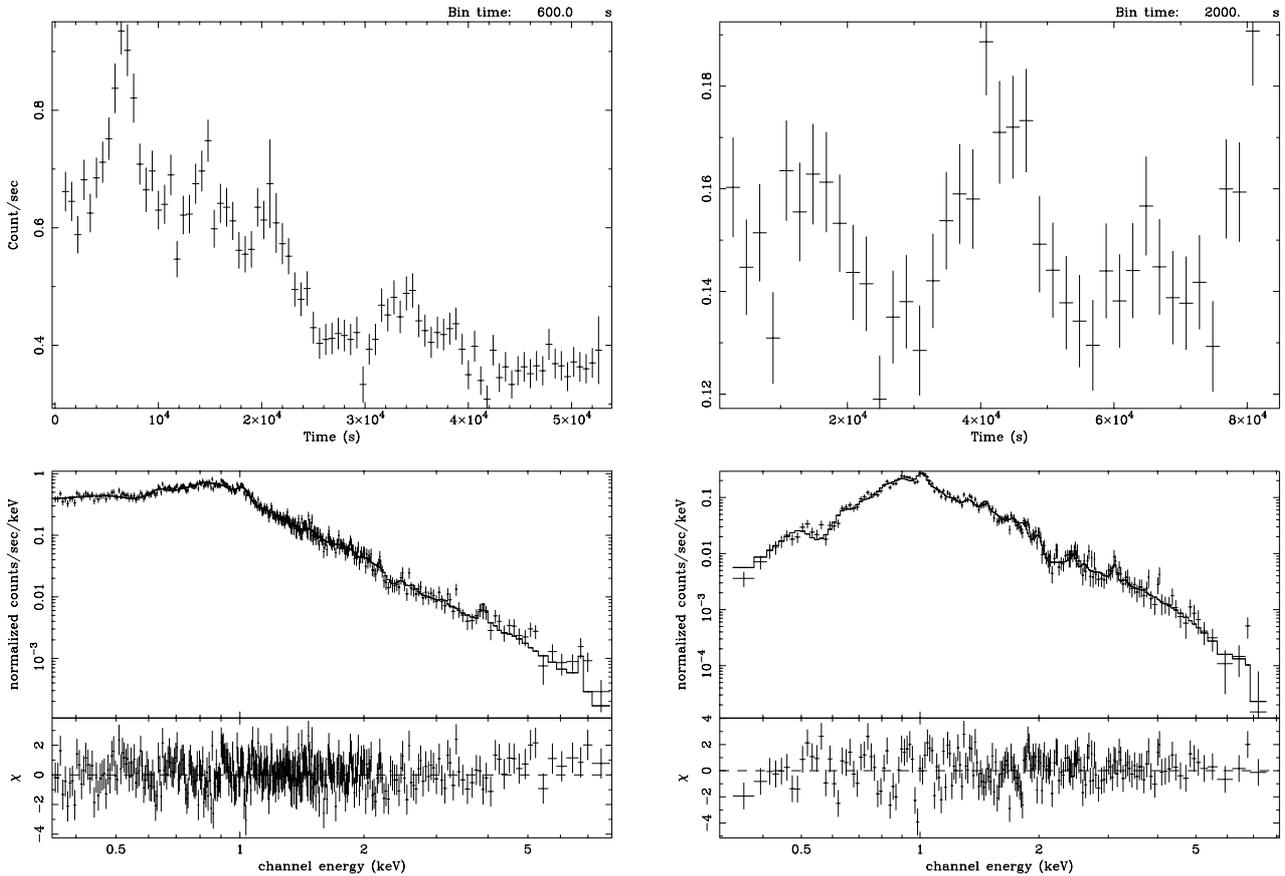

**Fig. 2.** Top: X-ray light curve of HD 285845 during the XMM-*Newton* and *Chandra* observations. Bottom: XMM-*Newton* and *Chandra* X-ray spectra of HD 285845. The fits to the spectra with absorbed 2-temperature plasma models with varying metal abundances are also shown.

The X-ray emission of source S26 does not show significant variability during our observations, and its X-ray spectrum is well described ($P = 0.93$) by a 2 temperature plasma model, with $kT_1 = 0.36 \pm 0.07$ keV, $kT_2 = 0.82 \pm 0.11$ keV (with comparable emission measure), $Z = 0.12 \pm 0.04$ and $N(\mathrm{H}) = (1.52 \pm 2.63) \times 10^{20}$ cm$^{-2}$. The low value for the hydrogen column density is consistent with the star being in front of the cloud.

In summary, of the 8 X-ray sources with a radio or optical counterpart which are not previously known pre-main sequence stars, 4 are stars, 3 of which unrelated to the L1551 cloud, while one (source S13) could be a pre-main sequence star belonging to the cloud and deserves further investigation. The other 3 are likely to be extra-galactic background sources and one is related to the proto-stellar jet HH 154.

### 3.1. Unidentified sources

We have analyzed the spectra of the remaining 10 X-ray bright sources with no optical or radio counterpart. Each of them is satisfactorily fit by an absorbed power law spectrum (although in many cases, as reported in Table 3, a thermal fit also provides an acceptable description to the spectrum), with best fit values summarized in Table 3. This, together to the absence of an optical counterpart in the USNO catalog indicates that these sources are most likely of extra-galactic origin.

Together with the 3 sources with optical or radio counterpart, 13 of the bright sources are thus likely to be extra-galactic. The presence of 13 extra-galactic serendipitous X-ray sources in the XMM-*Newton* field of view at a flux level of the order to $10^{-14}$ erg cm$^{-2}$ s$^{-1}$ is in rough agreement with the expected number density of background sources determined on the basis of the $\log N - \log S$ relationship for X-ray sources (see e.g. Hasinger et al., 2001), which predicts that at this flux limit 100–200 sources per square degree should be present in any given X-ray observation. The area covered by XMM-*Newton* field of view is approximately of 0.2 square deg, so that the expected number of serendipitous extra-galactic sources for a low absorption field is 20 to 40.

**Table 3.** Best fit spectral parameters for all the sources listed in Table 2 which are likely extra-galactic. The spectra were fit by an absorbed power-law model. $P$ is the probability level for the fit.

| Source | $N(H)$ $10^{22}$ cm$^{-2}$ | Spectral index $\gamma$ | $\chi^2$ | P | Notes |
|---|---|---|---|---|---|
| 4  | 0.49±0.17 | 1.7±0.3 | 1.25 | 0.25 | 1T fit equivalent (kT=8.2) – no counterpart |
| 5  | 0.23±0.14 | 2.6±0.7 | 1.14 | 0.30 | 1T fit equivalent ($kT = 2.2$) – no counterpart |
| 8  | 0.79±0.55 | 2.1±0.8 | 0.75 | 0.74 | 1T fit equivalent ($kT = 1.9$) – no counterpart |
| 9  | 0.58±0.31 | 3.7±1.3 | 0.52 | 0.92 | 1T fit equivalent ($kT = 0.9$) – no counterpart |
| 10 | 1.65±1.02 | 1.1±0.5 | 1.32 | 0.16 | 1T fit unacceptable – see text |
| 11 | 0.33±0.21 | 1.6±0.5 | 0.94 | 0.51 | 1T fit equivalent ($kT = 6.6$) – no counterpart |
| 12 | 0.14±0.16 | 1.2±0.4 | 0.72 | 0.70 | 1T fit equivalent ($kT = 44$) – radio source [SB86] L1551 3 |
| 15 | – | – | – | – | no fit possible, very hard spectrum – no counterpart |
| 16 | 1.89±1.24 | 1.6±0.7 | 0.92 | 0.55 | 1T fit unacceptable – radio source [SB86] L1551 2 |
| 17 | 0.30±0.16 | 1.0±0.2 | 0.86 | 0.70 | 1T fit equivalent ($kT = 64$) – no counterpart |
| 21 | 2.34±0.67 | 2.3±0.5 | 0.40 | 0.94 | 1T fit equivalent ($kT = 3.3$) – no counterpart |
| 24 | 4.26±3.07 | 3.1±1.8 | 2.40 | 0.02 | 1T fit equivalent ($kT = 2.3$) – no counterpart |
| 27 | 1.13±0.54 | 1.6±0.5 | 1.10 | 0.36 | 1T fit equivalent ($kT = 20$) – no counterpart |

## 4. X-ray sources associated with pre-main sequence stars in L1551

In this section we summarize our results on the previously known pre-main sequence stars that have been detected in the XMM-*Newton* observations of the L1551 star-forming complex. The EPIC-pn spectra of all these sources have been fit with absorbed two-temperature models; the fits are shown together with the spectra in Fig. 3. The best-fit model parameters are listed in Table 7. For the sources with enough statistics (and which showed visible systematic deviations in the fit residuals) spectral fits with two-temperature plasma models with varying individual metal abundance were performed. The procedure followed was to first fit the data with a variable $Z$ model, and then allow individual elements to vary individually, in order of atomic weight. If the best fit abundance for the given element was more than 1 $\sigma$ away from the one of Fe the element would be left free to vary, else it would be coupled back to Fe (which was always allowed to vary). The results of these fits are summarized in Table 8.

The light curves of the PMS X-ray sources are shown in Fig. 3. In order to evaluate the presence of X-ray variability, the Kolmogorov-Smirnov (K-S) test, which measures the maximum deviation of the integral photon arrival times from a constant source model, was applied. Table 9 summarizes the results of these statistical tests, providing also the source average flux during the XMM-*Newton* observations and the ROSAT and ASCA observations, as given by Carkner et al. (1996). The ROSAT and ASCA observations were spaced by 1 year. Below we discuss each star individually.

### 4.1. [BHS98] MHO 5

This late type star (M6) star is identified here with source S2, with a 1.4 arcsec offset between the optical and X-ray position. [BHS98] MHO 5 is identified as a member of the L1551 star-forming complex on the basis of the strong Li I $\lambda 6707$ feature in the optical spectrum (Briceño et al., 1998). The He I $\lambda 5876$ and the forbidden [O I] $\lambda 6300$ and $\lambda 6364$ lines in emission allow Briceño et al. (1998) to confirm [BHS98] MHO 5 as a CTTS, and the strong H$\alpha$ emission is indicative of high activity (Briceño et al., 1998).

The star was not detected in the ROSAT observations of the L1551 cloud (Carkner et al., 1996; Wichmann et al., 1996; König et al., 2001) and does not fall in the field of view of the *Chandra* observation (Bally et al., 2003).

The X-ray light curve shows an apparent increase of the X-ray count rate by approximately 50% over a 30 ks time. Nevertheless this variation is not statistically compelling, since the source has a probability of constancy greater than 50%.

The X-ray spectrum is well fit with an absorbed two-temperature model, with $kT_1 = 0.29 \pm 0.05$ keV and $kT_2 = 0.95 \pm 0.12$ keV and a low coronal metal abundance $Z = 0.19 \pm 0.12$.

### 4.2. V826 Tau

V826 Tau (identified here with source S3 with a 1.0 arcsec offset between X-ray and optical position) is a spectroscopic binary WTTS (Mundt et al., 1983) of K7 spectral type and a known member of the L1551 star-forming complex. X-ray emission from the star was first detected in *Einstein* X-ray data (Feigelson & Decampli, 1981; Reipurth et al., 1990), and seen again with ASCA and ROSAT by Carkner et al. (1996) who report of a large flare during the ASCA observations, with the source X-ray luminosity increasing by a factor $\sim 5$.

The X-ray emission from V826 Tau varies during the XMM-*Newton* observation, with its flux slightly increasing during 50 ks. This slow rise is statistically significant and the source probability of constancy is $\simeq 10^{-5}$. The source flux was $15.0 \times 10^{-13}$ erg cm$^{-2}$ s$^{-1}$ (0.2–2.0 keV) in the ROSAT observations and $15.8 \times 10^{-13}$ erg cm$^{-2}$ s$^{-1}$ (0.5–3.0 keV) in the ASCA one. We derive an average flux in the XMM-*Newton* observation of $\simeq 13 \times 10^{-13}$ erg cm$^{-2}$ s$^{-1}$ in both of the two above energy ranges, indicating lack of strong long term variability.

The X-ray spectrum of V826 Tau is not well described by the simple absorbed two-temperature model, with a reduced

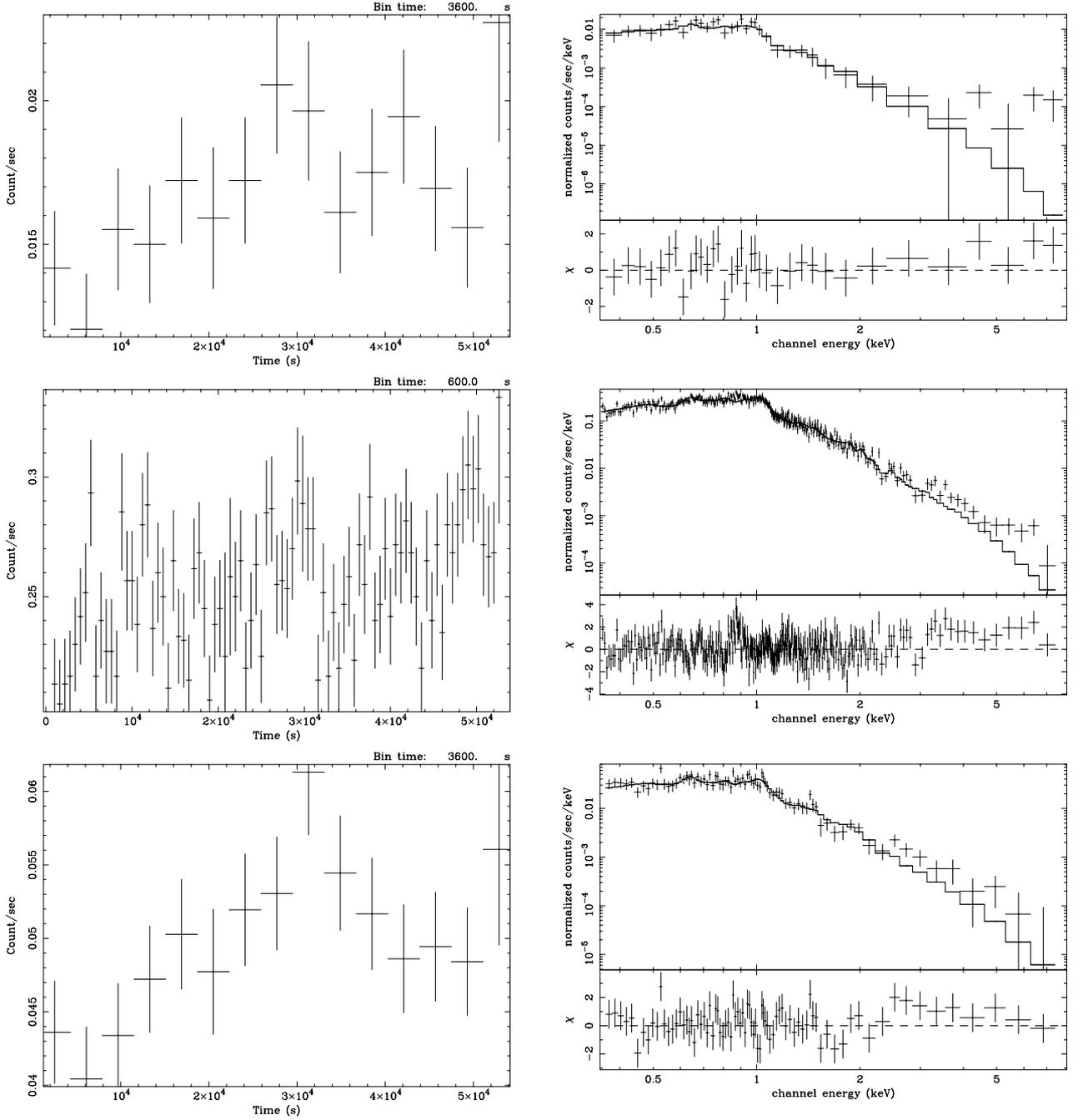

**Fig. 3.** EPIC-pn light curves and spectra for the pre-main sequence stars detected in the XMM-*Newton* observations of the star-forming complex L1551. Best fit 2-$T$ spectra are also plotted with the spectra. From top to bottom: [BHS98] MHO 5, V826 Tau, V827 Tau.

$\chi^2$ of the fit not acceptable ($P = 0.0015$). Indeed the residuals to the fit show significant structure, in particular around $E \simeq 0.9$ keV, where the Ne K$\alpha$ complex is located. Letting the abundance of some elements with strong lines free to vary individually improves the fit significantly ($P = 0.16$), as shown in Table 8. The resulting abundance values are Ne = $0.61 \pm 0.12$, Ca = $2.13 \pm 0.71$ and Ni = $0.61 \pm 0.31$, all significantly higher than the Fe metal abundance (Fe = $0.15 \pm 0.02$).

### 4.3. V827 Tau

V827 Tau (identified here with source S6, with a 8.9 arcsec offset) also is a known X-ray bright WTTS member of the L1551 star-forming complex of K7 spectral type. It was first detected in X-ray in *Einstein* data by Feigelson & Kriss (1981) and subsequently observed by ASCA and ROSAT (Carkner et al., 1996). The X-ray emission from the star appears to vary by about 50% over a $\sim 40$ ks time during the XMM-*Newton* observation, with a source probability of constancy of 0.04 according to the K-S test.

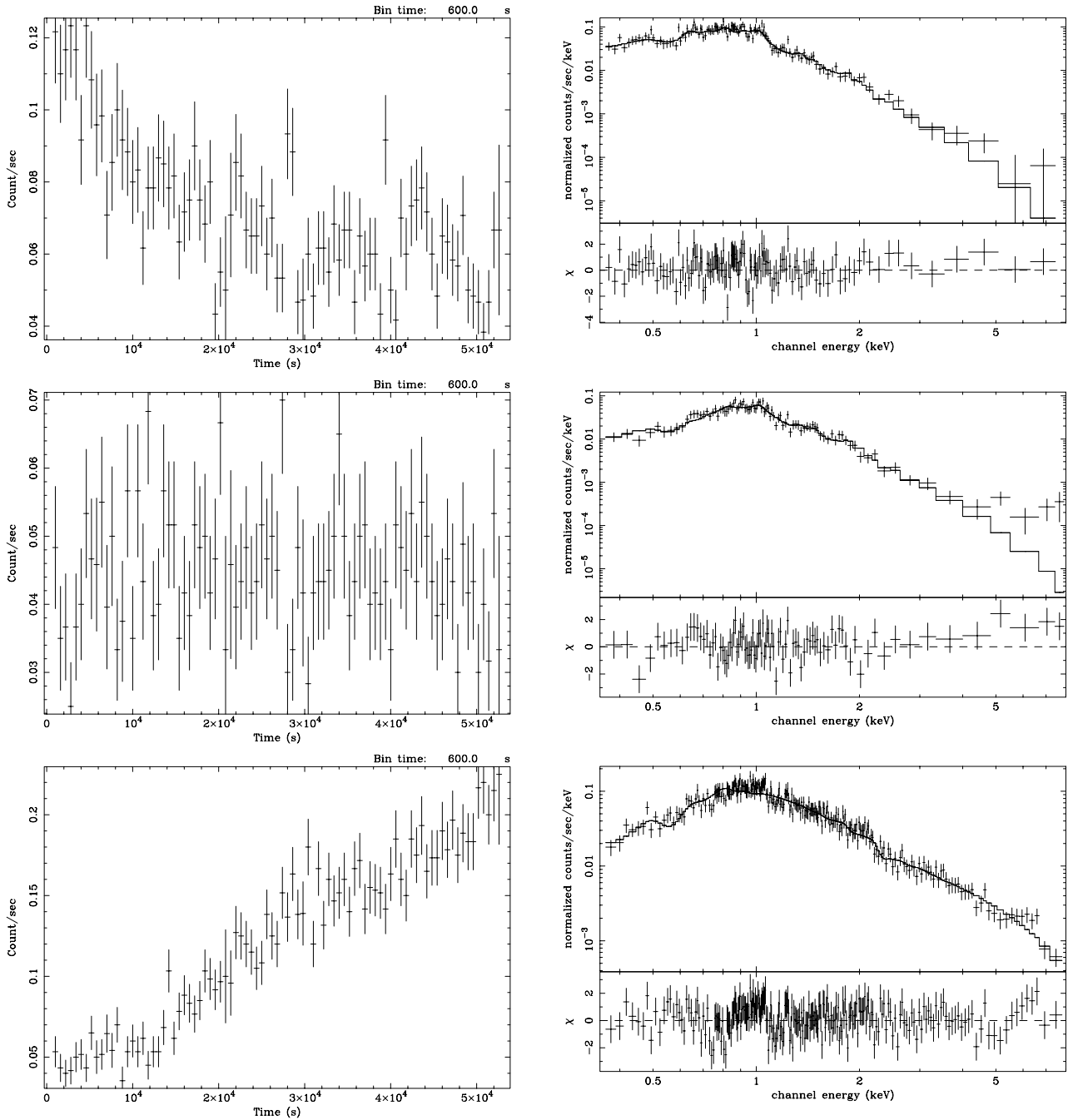

**Fig. 3.** *continued* – From top to bottom: V1075 Tau, V710 Tau, XZ Tau.

The ROSAT and ASCA fluxes estimate for V827 Tau reported by Carkner et al. (1996) are $13 \times 10^{-13}$ erg cm$^{-2}$ s$^{-1}$ (0.2–2.0 keV) and $4.5 \times 10^{-13}$ erg cm$^{-2}$ s$^{-1}$ (0.8–3.5 keV). The difference in the source flux between ROSAT and ASCA observations is likely to be due to the difference in energy range over which the flux is estimated; Carkner et al. (1996) compare the ROSAT and ASCA spectra and do not find any significant difference between the two. Over the same energy ranges used by Carkner et al. (1996), the fluxes derived from the XMM-*Newton* data are 1.9 and $1.5 \times 10^{-13}$ erg cm$^{-2}$ s$^{-1}$, respectively, significantly lower than the values found by Carkner et al. (1996), indicating long term variability of the X-ray emission from V827 Tau.

The X-ray spectrum is well described by an absorbed two-temperature model, with $kT_1 = 0.30 \pm 0.03$ keV and $kT_2 = 1.21 \pm 0.08$ keV and a low coronal metal abundance of $Z = 0.18 \pm 0.06$. Since the spectrum has good $S/N$ a fit with a variable abundance 2-temperature plasma model was also performed, finding S and Ni to be significantly over-abundant with respect to the other elements (Table 8).

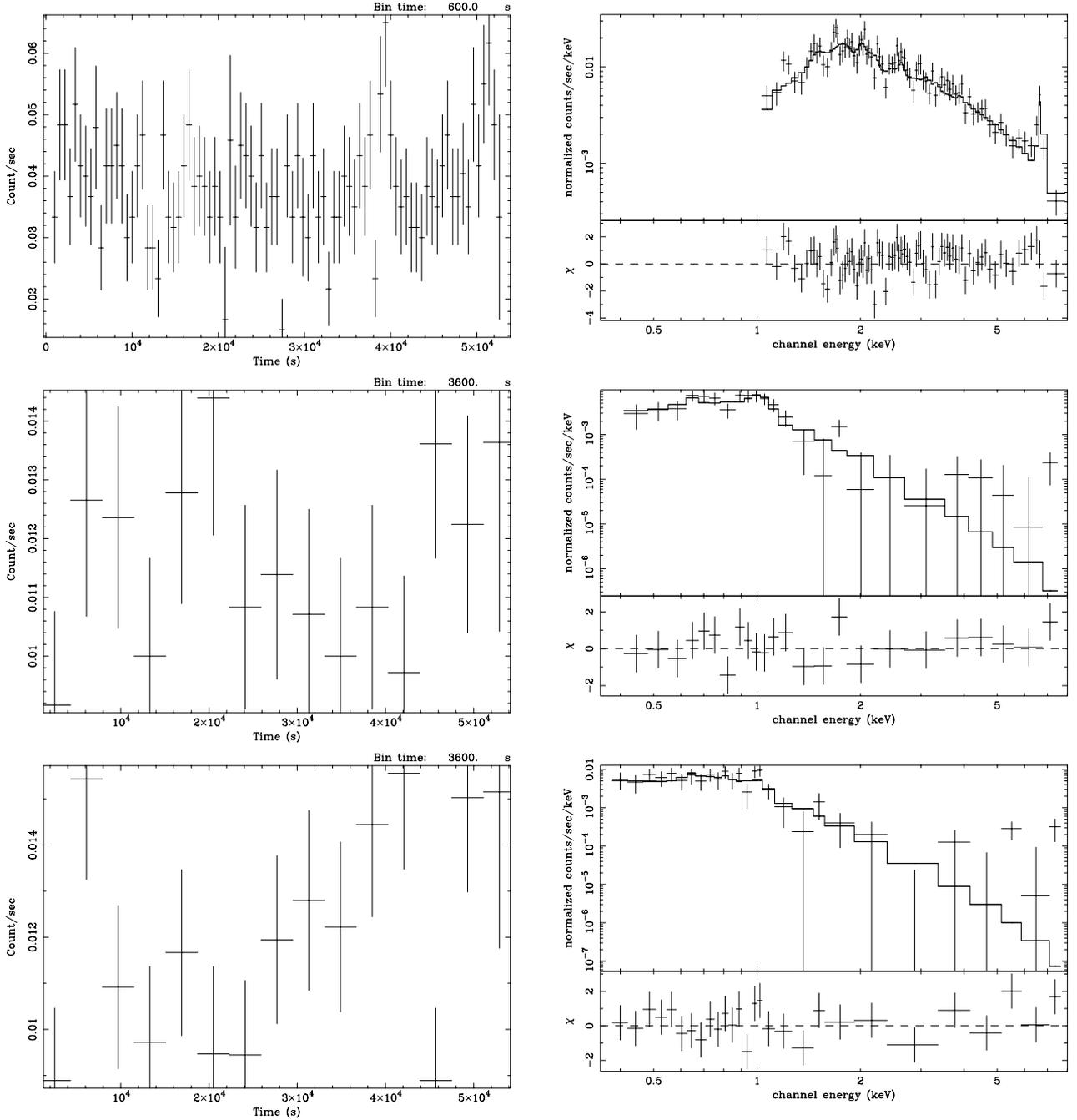

**Fig. 3.** *continued* – From top to bottom: HL Tau, [BHS98] MHO 4, [BHS98] MHO 9.

### 4.4. V1075 Tau

V1075 Tau (identified here with source S7 with a 1.8 arcsec offset) is a binary WTTS of spectral type K7, and a known member of the L1551 star-forming complex. Its bright X-ray emission was first detected in *Einstein* X-ray data by Feigelson & Kriss (1981) and it has been subsequently observed by ASCA and ROSAT (Carkner et al., 1996).

The average spectrum of the source is well described by an absorbed two-temperature model, with $kT_1 = 0.35 \pm 0.03$ keV and $kT_2 = 1.03 \pm 0.04$ keV and a low coronal metal abundance $Z = 0.15 \pm 0.03$. The fit however shows significant residuals around $E \sim 0.9$ keV likely to be due to an overabundance of Ne in the star's corona. A variable abundance fit indeed shows that a higher Ne abundance of $0.44 \pm 0.16$ improves the fit, eliminating the structure in the residuals around 0.9 keV.

The temporal variability of V1075 Tau during the XMM-*Newton* observation is significant, with the source counts decreasing by a factor of $\sim 2$ in about 30 ks. The good statistics of the light curve allowed the spectrum to be analyzed during 2 phases of the observations, which was split into the first 30 ks and the last 24 ks. The two spectra are showed in Fig. 4. While the temperatures in the spectrum do not change significantly, the change in the source flux is dominated by a decrease in the emission measure for the cooler temperature. Also, a higher absorption is seen when the source flux is more intense. The best-fit parameters for the two fits are summarized in Table 4.

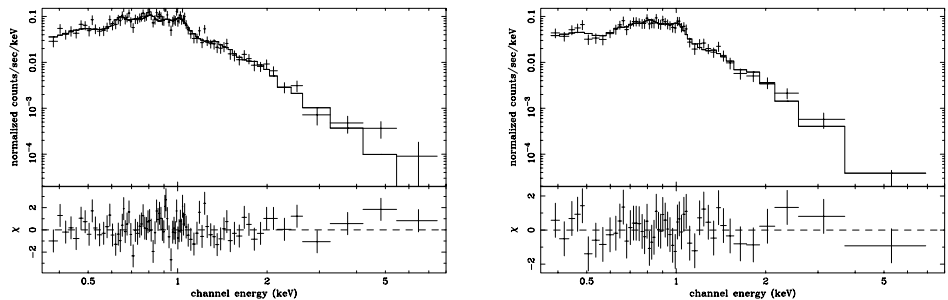

**Fig. 4.** Left: the spectrum of V1075 Tau from the first 30 ks of observation; right: the spectrum from the last 24 ks. The fits to the spectra with an absorbed 2-$T$ model are also shown.

**Table 4.** Best-fit spectral parameters for V1075 Tau during two consecutive time intervals of XMM-*Newton* observations.

| Time interval | $N(\mathrm{H})$ | $kT_1$ | $kT_2$ | $EM_1$ | $EM_2$ | $Z$ | $\chi^2$ | $P$ |
|---|---|---|---|---|---|---|---|---|
| ks | $10^{22}$ cm$^{-2}$ | keV | keV | $10^{53}$ cm$^{-6}$ | $10^{53}$ cm$^{-6}$ | $Z_\odot$ | | |
| 0–30 | $0.19 \pm 0.03$ | $0.32 \pm 0.02$ | $1.11 \pm 0.9$ | $3.18 \pm 2.61$ | $1.84 \pm 0.57$ | $0.15 \pm 0.05$ | 1.19 | 0.11 |
| 30–54 | $0.08 \pm 0.03$ | $0.47 \pm 0.11$ | $1.03 \pm 0.08$ | $0.73 \pm 0.57$ | $0.93 \pm 0.52$ | $0.19 \pm 0.06$ | 0.68 | 0.96 |

The flux levels in ASCA and ROSAT data are similar: $4.7 \times 10^{-13}$ erg cm$^{-2}$ s$^{-1}$ (0.5–3.0 keV) for ASCA and $5.3 \times 10^{-13}$ erg cm$^{-2}$ s$^{-1}$, (0.2–2.0) keV for ROSAT. In these two energy ranges the average flux in the XMM-*Newton* observation is $4.1 \times 10^{-13}$ erg cm$^{-2}$ s$^{-1}$ and $4.3 \times 10^{-13}$ erg cm$^{-2}$ s$^{-1}$, respectively, indicating a lack of significant long term variability.

### 4.5. V710 Tau A&B

V710 Tau (identified with source S14 with a 3.7 arcsec offset) is a binary system composed of a CTTS and a WTTS (Carkner et al., 1996) and a known X-ray luminous member of the star-forming complex with spectral types M1 and M3. X-ray emission from V710 Tau was detected in *Einstein* data (Damiani et al., 1995) and it has been subsequently observed with ROSAT (Carkner et al., 1996). The two stars are not resolved in the XMM-*Newton* observations. V710 Tau is not seen to vary during ROSAT observations, and it does not show significant variability in our data.

The X-ray spectrum is well fit by an absorbed two-temperature model, with $kT_1 = 0.63 \pm 0.05$ keV and $kT_2 = 1.24 \pm 0.10$ keV and a coronal metal abundance $Z = 0.22 \pm 0.07$. The fit can be improved by letting the relative abundance of metals free to vary individually. This results in a high Ne abundance of $0.98 \pm 0.45$, with a change in the model null hypothesis probability from 0.37 to 0.45.

### 4.6. XZ Tau

XZ Tau (identified here with source S18 with a 1.4 arcsec offset) is a well known M3 CTTS belonging to L1551 star-forming complex. The star has been resolved as a binary in the infrared, with a 0.3 arcsec separation (Haas et al., 1990). It is associated, together with HL Tau, with a complex set of bipolar jets and Herbig-Haro outflows (Mundt et al., 1990). X-ray emission from XZ Tau was detected in *Einstein* data (Damiani et al., 1995) and it has been subsequently observed with ROSAT and ASCA (Carkner et al., 1996). In the XMM-*Newton* observations the X-ray emission from this source is for the first time clearly resolved from the nearby HL Tau (discussed below).

The average spectrum of XZ Tau is well described by an absorbed two-temperature model with $kT_1 = 0.23 \pm 0.02$ keV and $kT_2 = 3.57 \pm 0.33$ keV and a very low coronal metal abundance of $Z = 0.007 \pm 0.004$. Allowing individual abundances to vary does not provide a significantly better description to the time integrated spectral data.

The short-term temporal variability of XZ Tau during the XMM-*Newton* observation is remarkable: its X-ray count rate increases in an approximately linear fashion along the duration of the observation, brightening by a factor of $\simeq 4$ in 50 ks (see Fig. 3). Given the significant temporal variability and the statistics of the XMM-*Newton* data we have also performed a time-resolved spectral analysis, dividing the observation in three segments of 20, 20 and $\simeq 15$ ks each. The three spectra are shown in Fig. 5, and their best-fit parameters are listed in Table 5. The hydrogen column density decreases, during the XMM observation, from $1.06 \times 10^{22}$ cm$^{-2}$ to $0.26 \times 10^{22}$ cm$^{-2}$, while the temperatures of the two components increases from $kT_1 = 0.14$ keV and $kT_2 = 2.29$ keV to $kT_1 = 1.00$ keV and $kT_2 = 4.98$ keV. The metal abundance remains constantly low, and the intrinsic variability is driven by the evolution of the emission measure of the cooler component.

Although XZ Tau is not resolved from the nearby HL Tau in the ROSAT observations nor in the ASCA ones, the amount of contamination in the determination of the flux is likely to be negligible in the ROSAT data, given the very absorbed spec-trum of HL Tau, which results in a very small flux in the soft ROSAT passband. In fact, HL Tau is not visible even in the ROSAT HRI data, which have sufficient spatial resolution to in principle resolve it.

The contamination could be more sizable in the ASCA spectra, given that HL Tau has a very hard spectrum. However, Carkner et al. (1996) quote the ASCA flux in the same energy band as the ROSAT one (0.5–2.0 keV), where, again, the contribution from HL Tau is likely to be negligible.

### 4.6.1. *Chandra* data

During the *Chandra* observation the temporal behavior of XZ Tau is unremarkable, with a quite constant light curve (Fig. 6). The spectral analysis of the ACIS data shows the source to be well fit by a simple 2-temperature model with parameters $kT = 0.65 \pm 0.03, 1.56 \pm 0.12$ keV, and comparable emission measure for the two components. The flux in the 0.5–2.0 keV range is $0.96 \times 10^{-14}$, about 3 times lower than during the XMM-*Newton* observation (although the problem with large absorption in the *Chandra* data likely makes the flux somewhat underestimated), and comparable to the flux reported by Carkner et al. (1996) for the ROSAT PSPC observation. During the *Chandra* observation XZ Tau is softer than during the XMM-*Newton* one, and the best-fit metal abundance appears to be higher, at $Z = 0.3 \pm 0.09$; whether this is a real effect, or partly due to the different characteristics (and calibration issues) of the *Chandra* and XMM-*Newton* detectors, is not clear. Nevertheless, the good agreement between the metal abundance determined with the *Chandra* and XMM-*Newton* data for HD 285845 and JH 188 would point to the changes in metal abundance in the XZ Tau spectrum to be real.

### 4.6.2. Long-term variability of XZ Tau

In addition to the short term variability visible in the XMM-*Newton* observation, XZ Tau exhibits also a significant long term variability. Carkner et al. (1996) report a variation of a factor $\simeq 15$ between their ROSAT and ASCA observations (which includes the contribution from the nearby and unresolved HL Tau, Sect. 4.6), with source flux increasing from $\simeq 10^{-13}$ to $\simeq 15 \times 10^{-13}$ erg cm$^{-2}$ s$^{-1}$ over the energy range 0.5–2.0 keV. During the XMM-*Newton* observation, over the same energy range, the average source flux of $2.9 \times 10^{-13}$ erg cm$^{-2}$ s$^{-1}$ is about three times the ROSAT level, while during the *Chandra* observation the source flux $0.96 \times 10^{-13}$ erg cm$^{-2}$ s$^{-1}$ is again close to the level observed by ROSAT.

Given this large variability and the significant difference between the XMM-*Newton* and *Chandra* observations discussed here, we also examined the long-term behavior of XZ Tau by looking at the other ROSAT (PSPC and HRI) observations not already reported in the literature as well as at the *Einstein* observation. The results are summarized in Table 6. With the exception of the ASCA observation the X-ray flux from XZ Tau spans a factor of $\simeq 3$. During the ASCA observation the star was in an unusual state, which does not appear to be a flare, given that its light curve was very flat and featureless Carkner et al. (1996).

**Table 6.** Count rate (in counts per second) and flux level in the 0.5–2.0 keV band for all known X-ray observations of XZ Tau. Flux in units of $10^{-13}$ erg cm$^{-2}$ s$^{-1}$.

| Date | Instr. | Rate | Flux |
|---|---|---|---|
| 1981-31-01 | IPC | 0.017 | 1.82 |
| 1994-23-02 | ASCA (SIS) | 0.06 | 15 |
| 1996-12-31 | PSPC | 0.023 | 1.13 |
| 1997-02-10 | PSPC | 0.021 | 1.05 |
| 1997-05-31 | HRI | 0.0132 | 1.69 |
| 1999-12-14 | HRI | 0.0162 | 2.07 |
| 2000-09-09 | XMM (pn) | 0.13 | 2.9 |
| 2001-07-23 | Chandra | 0.024 | 0.96 |

### 4.7. HL Tau

HL Tau (identified here with source S19 with a 2.3 arcsec offset) lies about 23 arcsec from XZ Tau and its X-ray emission has not been resolved from its brighter nearby companion before. HL Tau is an embedded young stellar object (spectral type K7) often considered a prototype very young low-mass star, with a circumstellar disk that resembles the solar nebula at the early stages of planet formation (Men'shchikov et al., 1999). Together with XZ Tau it is associated with bipolar jets and Herbig-Haro outflows (Mundt et al., 1990). It does not show significant variability during the XMM-*Newton* observation.

HL Tau is visible in the XMM-*Newton* image only at energies above 1 keV, and therefore we analyzed the source spectrum in the energy range 1.0–7.5 keV. The spectrum is well described by an absorbed thermal model with a single temperature component with high absorption, $N(\mathrm{H})= 2.05 \times 10^{22}$ cm$^{-2}$, a temperature $kT = 2.9 \pm 0.2$ keV and a metal abundance $Z = 0.56 \pm 0.14$.

### 4.7.1. *Chandra* observation

The light curve of the *Chandra* observation of HL Tau shows evidence for a small short duration flare. The ACIS spectrum of the source has very similar spectral parameters as the EPIC pn one, with $T = 2.86 \pm 0.26$ keV and a coronal metal abundance of $Z = 0.22 \pm 0.08$. The metal abundance is nominally lower in the *Chandra* data than in the XMM-*Newton* ones, and indeed the strong Fe K line visible in the EPIC data is not prominent in the ACIS ones; whether this difference is significant, given the two different instruments involved, is not clear.

### 4.8. [BHS98] MHO 4

[BHS98] MHO 4 is identified with source S23 with a 3.2 arcsec offset. The spectrum of this star obtained by Briceño et al. (1998) shows Li I $\lambda$6707 strongly in absorption. He I $\lambda$5876, [O I] $\lambda$6300 and [O I] $\lambda$6363 are also detected, confirming this very late type star (M6) as a CTTS. The star appears located in

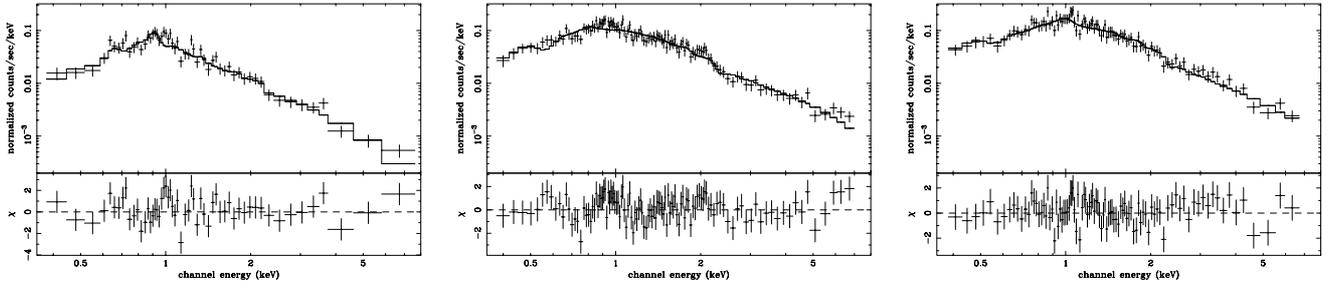

**Fig. 5.** The spectra of XZ Tau during three different intervals of the XMM-*Newton* observation. From left to right: the spectrum from the first 20 ks, the spectrum from the subsequent 20 ks and the spectrum from the last 14 ks. The fits to the spectra with an absorbed 2-*T* model are also shown.

**Table 5.** Best-fit spectral parameters for XZ Tau during three consecutive time intervals of XMM-*Newton* observation.

| Time interval | $N(H)$ | $kT_1$ | $kT_2$ | $EM_1$ | $EM_2$ | $Z$ | $\chi^2$ | $P$ |
|---|---|---|---|---|---|---|---|---|
| ks | $10^{22}$ cm$^{-2}$ | keV | keV | $10^{53}$ cm$^{-6}$ | $10^{53}$ cm$^{-6}$ | $Z_\odot$ | | |
| 0–20 | $1.06 \pm 0.06$ | $0.14 \pm 0.08$ | $2.29 \pm 0.30$ | $1665 \pm 1297$ | $1.25 \pm 0.36$ | $0.057 \pm 0.035$ | 1.43 | 0.02 |
| 20–40 | $0.49 \pm 0.11$ | $0.29 \pm 0.07$ | $4.27 \pm 0.78$ | $36.7 \pm 44.3$ | $1.97 \pm 0.57$ | $0.0077 \pm 0.0076$ | 1.00 | 0.49 |
| 40–54 | $0.26 \pm 0.02$ | $1.00 \pm 0.08$ | $4.98 \pm 2.15$ | $2.12 \pm 2.89$ | $1.85 \pm 1.13$ | $0.083 \pm 0.066$ | 0.94 | 0.64 |

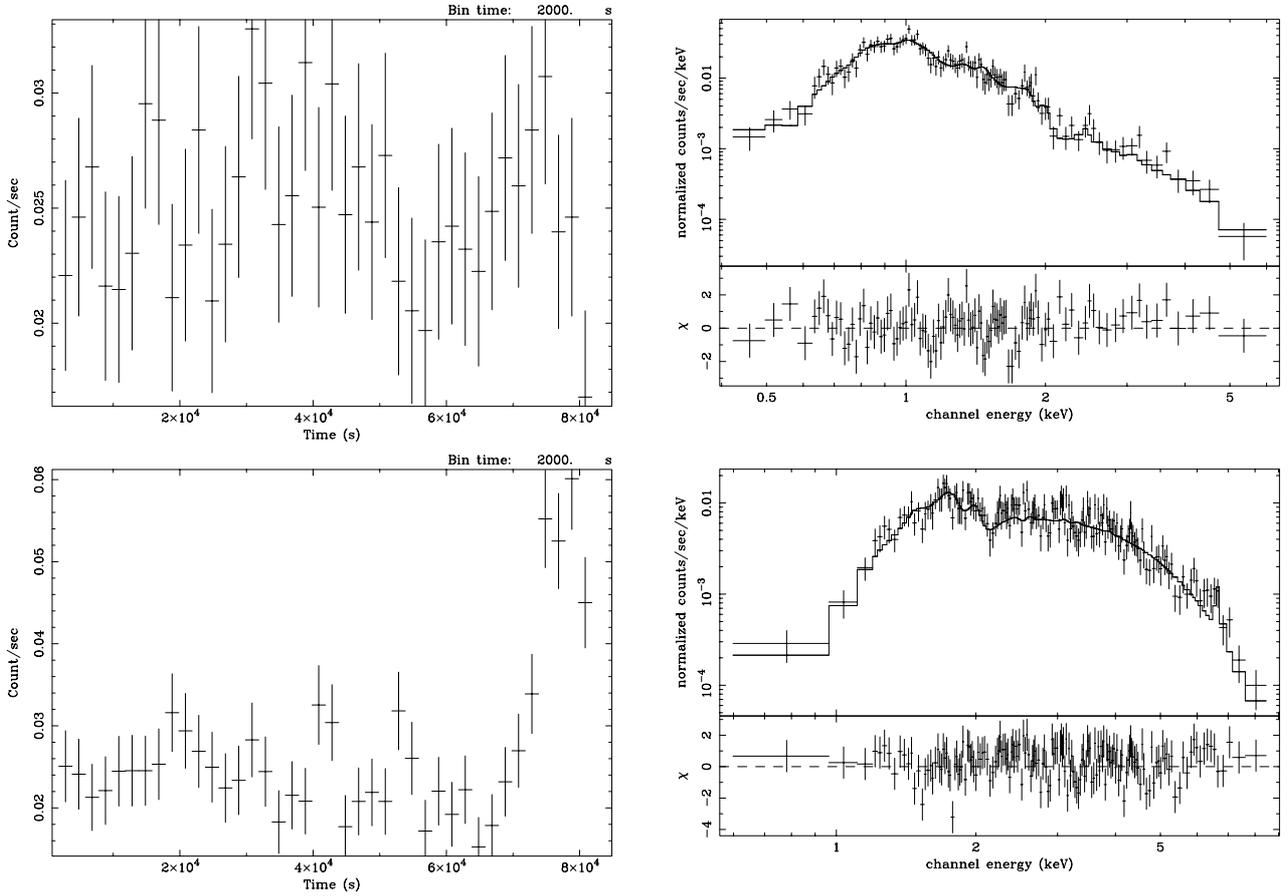

**Fig. 6.** *Chandra* ACIS light curves and spectra (together with best-fitting 2-*T* spectra) of (from top to bottom): XZ Tau, HL Tau.

a high extinction region about 8 arcmin south of L1551 IRS 5 (Briceño et al., 1998).

X-ray emission from this star was first detected by Carkner et al. (1996) in ROSAT observations of the field. Although they did not have the study by Briceño et al. (1998) available they correctly identified it as a new X-ray emitting T Tauri star. The source flux does not appear to have changed significantly since the ROSAT observation. [BHS98] MHO 4 also does not show significant variability during XMM-*Newton* observations.

The X-ray spectrum is well described by an absorbed two-temperature model, with $kT_1 = 0.33 \pm 0.09$ keV and $kT_2 = 1.23 \pm 0.15$ keV and a coronal metal abundance only loosely constrained by the upper limit $Z < 2.92$. The spectral fit does not strongly constrains the hydrogen column density, $N(\text{H}) \lesssim 2 \times 10^{21}$ cm$^{-2}$.

### 4.9. [BHS98] MHO 9

[BHS98] MHO 9 (identified here with source S25 with a 2.5 arcsec offset) is an other of the new WTTS cataloged by Briceño et al. (1998) in L1551, who attributed it spectral type M4.

X-ray emission from [BHS98] MHO 9 was not detected in the observations of L1551 made by ROSAT (Carkner et al., 1996; Wichmann et al., 1996; König et al., 2001). During the XMM-*Newton* observation the star does not exhibit significant variability. Its X-ray spectrum is also well described with an absorbed two-temperature model with $kT_1 = 0.34 \pm 0.07$ and $kT_2 = 1.11 \pm 0.41$ and a coronal metal abundance only loosely constrained by the upper limit $Z < 0.97$. The hydrogen column density is also loosely constrained, $N(\text{H}) \lesssim 2 \times 10^{21}$.

## 5. Discussion

Most known pre-main sequence stars belonging to the L1551 cloud and falling within the XMM-*Newton* field of view have been detected in the 50 ks observation discussed in the present paper, so that our survey can be considered to be nearly complete. X-ray emission from two of the new pre-main sequence stars discovered in the region by Briceño et al. (1998), MHO 5 and MHO 9, has been detected for the first time, bringing to nine the total number of T Tauri stars belonging to the cloud for which optical and X-ray data are available. Four of these (V826 Tau, V1075 Tau, V710 Tau and XZ Tau) are known binaries. Thanks to the large collecting area of XMM-*Newton* we were able to derive and analyze CCD-resolution X-ray spectra and light curves for all of these 9 pre-main sequence stars. While this sample of T Tau stars is too small to carry out a statistical study of correlations between X-ray emission and stellar properties, several interesting individual properties of their X-ray emission have been uncovered by the observation discussed here.

Few of the known pre-main sequence stars in the region are not detected, in particular the parent star of the HH 30 bipolar jet; given the "edge-on" position of the disk, very well visible in e.g. HST images (Burrows et al., 1996), it is possible that intrinsic X-ray emission is effectively shadowed by the disk itself.

### 5.1. Metal abundance

The target stars are all very young, and have recently formed from the same cloud, so that their photospheric abundances are likely to be very similar. At the same time, their coronal metal abundances deduced from the analysis of their X-ray spectra show a broad range of values. The metal abundance in the coronae of the WTTSs is clustered around $Z \simeq 0.2$ with a narrow spread, while the abundance derived for the CTTSs systems spans a much wider range ($Z \simeq 0.007$ to $Z \simeq 0.6$, i.e. almost two orders of magnitude).

In the cases in which the X-ray spectra have sufficient statistics, individual metal abundances were derived, showing that a number of elements tend to be consistently enhanced over Fe (which normally dominates, with the large number of Fe L lines, the determination of metal abundance from CCD-resolution spectra). This is in particular true for Ne, which shows, in V826 Tau, V1075 Tau and V710 Tau a consistent enhancement over Fe of a factor of $\simeq 3$ to $4$. All three of these stars are binary systems, with V826 Tau and V1075 Tau being WTTS, and V710 Tau having a WTTS and a CTTS component. No Ne enhancement is seen in the CTTSs in the sample. The same Ne enhancement is found in HD 285845, which is an active binary system past the PMS stage. This general enhancement of Ne over Fe is consistent with the coronal enhancement of noble gases found in high resolution spectra in active binaries (Drake et al., 2001).

Whatever the mechanism causing such a large range of coronal metal abundances starting from the same photospheric abundance, it appears to consistently enhance some elemental abundances in the WTTS systems (for which the X-ray emission mechanism is likely to be similar to the one in older, high-activity stars). Also, in some cases the coronal abundance appear to vary in time, as shown by the different abundances derived from the XMM-*Newton* and *Chandra* spectra for XZ Tau.

### 5.2. Temporal variability

The high temporal variability of X-ray emission from very young stars is known since the first *Einstein* observations (e.g. Montmerle et al., 1983), however the limited collecting area and spectral resolution of the previous generation of X-ray telescopes did not allow to study this temporal variability in detail. In our sample two sources show significant variability over short time scales: XZ Tau and V1075 Tau. For these we are able to derive spectral information during different phases of their light curve.

The X-ray flux of XZ Tau, a particularly interesting source because of its extreme young age (estimated at 55 000 yr, Beckwith et al., 1990), is seen to increase by a factor $\simeq 4$ during the 50 ks of XMM-*Newton* observations, while staying nearly constant during the 80 ks *Chandra* observation one year later. During the XMM-*Newton* observation the spectrum of the source also changes, with the absorbing column density decreasing by a factor $\simeq 4$ and hotter plasma appearing, with the high temperature going from $\simeq 2$ keV to $\simeq 5$ keV. It is difficult to explain the flux variation in XMM-*Newton* data with a stellar flare, which would normally present a faster and more impulsive rise phase, with a faster increase in plasma temperature.

**Table 7.** Best-fit spectral parameters for the X-ray-bright pre-main sequence stars detected in the XMM-*Newton* EPIC data. All spectra have been fit with an absorbed 2 temperature plasma model in the energy range 0.35–7.5 keV, apart from HL Tau whose spectrum has been fit with an absorbed 1 temperature plasma model over the energy range 1.0–7.5 keV (see text). In the column "type", b indicates that the star is a known binary system, W stands for WTTS and C for CTTS.

| Src | Name | Type | $N(H)$ | $kT_1$ | $kT_2$ | $EM_1$ | $EM_2$ | $Z$ | $\chi^2$ | $P$ |
|---|---|---|---|---|---|---|---|---|---|---|
| | | | $10^{22}$ cm$^{-2}$ | keV | keV | $10^{53}$ cm$^{-6}$ | $10^{53}$ cm$^{-6}$ | $Z_\odot$ | | |
| S2 | MHO 5 | C | $0.07 \pm 0.06$ | $0.29 \pm 0.05$ | $0.95 \pm 0.12$ | $0.10 \pm 0.18$ | $0.10 \pm 0.09$ | $0.19 \pm 0.12$ | 0.83 | 0.73 |
| S3 | V826 Tau | bW | $0.10 \pm 0.09$ | $0.33 \pm 0.01$ | $1.15 \pm 0.03$ | $3.74 \pm 1.04$ | $4.74 \pm 0.50$ | $0.17 \pm 0.02$ | 1.27 | 0.0015 |
| S6 | V827 Tau | W | $0.07 \pm 0.02$ | $0.30 \pm 0.03$ | $1.21 \pm 0.08$ | $0.41 \pm 0.37$ | $0.64 \pm 0.21$ | $0.18 \pm 0.06$ | 1.06 | 0.24 |
| S7 | V1075 Tau | bW | $0.15 \pm 0.02$ | $0.35 \pm 0.03$ | $1.03 \pm 0.04$ | $1.90 \pm 0.97$ | $1.58 \pm 0.35$ | $0.15 \pm 0.03$ | 0.97 | 0.57 |
| S14 | V710 Tau | bCW | $0.25 \pm 0.03$ | $0.63 \pm 0.05$ | $1.24 \pm 0.10$ | $0.65 \pm 0.38$ | $0.93 \pm 0.33$ | $0.22 \pm 0.067$ | 1.04 | 0.37 |
| S18 | XZ Tau | bC | $0.66 \pm 0.05$ | $0.23 \pm 0.02$ | $3.57 \pm 0.33$ | $141 \pm 78$ | $1.86 \pm 0.28$ | $0.007 \pm 0.004$ | 1.05 | 0.29 |
| S19 | HL Tau | C | $2.05 \pm 0.12$ | $2.92 \pm 0.23$ | – | $1.80 \pm 0.29$ | – | $0.56 \pm 0.14$ | 1.19 | 0.12 |
| S23 | MHO 4 | C | $\le 0.15$ | $0.33 \pm 0.09$ | $1.23 \pm 0.15$ | $1.16 \pm 1.76$ | $0.008 \pm 0.03$ | $0.02 \pm 0.06$ | 0.85 | 0.64 |
| S25 | MHO 9 | W | $\le 0.15$ | $0.34 \pm 0.07$ | $1.11 \pm 0.41$ | $0.034 \pm 0.16$ | $0.27 \pm 0.064$ | $0.34 \pm 0.63$ | 1.02 | 0.43 |

**Table 8.** Best-fit spectral parameters for the X-ray sources for which variable abundance fits were performed.

| Source | S3 (V826 Tau) | S6 (V827 Tau) | S7 (V1075 Tau) | S14 (V710 Tau) | S19 (HL Tau) | S22 (HD 285845) |
|---|---|---|---|---|---|---|
| $N(H)$ | $0.096 \pm 0.010$ | $0.08 \pm 0.04$ | $0.14 \pm 0.03$ | $0.23 \pm 0.03$ | $1.93 \pm 0.14$ | $0.032 \pm 0.007$ |
| $kT_1$ | $0.36 \pm 0.02$ | $0.30 \pm 0.03$ | $0.38 \pm 0.02$ | $0.66 \pm 0.04$ | $3.16 \pm 0.30$ | $0.67 \pm 0.01$ |
| $kT_2$ | $1.36 \pm 0.06$ | $1.19 \pm 0.10$ | $1.30 \pm 0.13$ | $1.56 \pm 0.18$ | – | $2.06 \pm 0.76$ |
| $\chi^2$ | 1.08 | 0.97 | 0.86 | 1.01 | 1.15 | 1.04 |
| $P$ | 0.16 | 0.56 | 0.86 | 0.45 | 0.16 | 0.31 |
| N | = Fe | = Fe | = Fe | = Fe | = Fe | = Fe |
| O | = Fe | = Fe | = Fe | = Fe | = Fe | $0.42 \pm 0.08$ |
| Ne | $0.61 \pm 0.12$ | = Fe | $0.44 \pm 0.16$ | $0.98 \pm 0.45$ | = Fe | $1.11 \pm 0.15$ |
| Mg | = Fe | = Fe | = Fe | = Fe | = Fe | = Fe |
| Si | = Fe | = Fe | = Fe | = Fe | = Fe | $0.13 \pm 0.06$ |
| S | = Fe | $1.71 \pm 0.67$ | = Fe | = Fe | $1.453 \pm 0.61$ | = Fe |
| Ca | $2.13 \pm 0.71$ | = Fe | = Fe | = Fe | = Fe | $2.37 \pm 0.63$ |
| Fe | $0.15 \pm 0.02$ | $0.18 \pm 0.09$ | $0.11 \pm 0.04$ | $0.27 \pm 0.11$ | $0.49 \pm 0.13$ | $0.26 \pm 0.04$ |
| Ni | $0.61 \pm 0.31$ | $1.38 \pm 1.01$ | = Fe | = Fe | = Fe | = Fe |

Most likely, the decrease in absorbing column density as the flux increases is not physically linked to the emitting region; its inverse correlation with the total count rate points to the variation being due, at least in part, to a "shadow" effect from material passing in front of the source. As reported by Carkner et al. (1996), at the time of the ASCA observation, when the source was at its brightest, the best-fit absorbing column density (at $N(H) = 1.6 \times 10^{21}$) was also lower than during the XMM-*Newton* observation. In parallel with this possible shadow effect significant variations in the intrinsic X-ray emission can of course take place. Unfortunately, the *Chandra* data, given the impossibility of accurately determining the absorbing column density, do not allow to test the presence of a correlation between apparent source luminosity and absorbing column density in XZ Tau.

One possible explanation of the spectral variation observed in XZ Tau is that one is actually observing the shadowing from the stream of accreting material along the star magnetic field lines, ending in the hot accretion spot on the star's photosphere. The value for peak column density determined from the XMM observations is compatible with the column density expected for the accretion stream in low-mass CTTS. In this case most of the X-ray emission, to be effectively eclipsed, should originate from the accretion spot itself, and thus be accretion driven, rather than coronal in origin. In this case, rotational modulation of the X-ray emission should also be present, if the accretion spot(s) are self-eclipsed by the star itself.

## 6. Conclusions

The high $S/N$ ratio afforded by the XMM-*Newton* EPIC observations discussed here allows to study the emission characteristics of the target population at a level of detail not possible before. Although the number of pre-main sequence targets with high statistics is small, significant differences are visible in the X-ray emission of the WTTS and CTTS population. The metal abundance of the plasma responsible for the X-ray emission of the two groups shows a different behavior, with the WTTS clustering tightly around $Z \simeq 0.2$, and showing the same Ne overabundance which is seen in the spectra of very active stars. At the same time the CTTS span a much wider range in $Z$, and show no evidence for Ne overabundance. Thus, while the X-ray spectra of the WTTS population are indistinguishable from the ones of older highly active stars (such as active binaries), the spectra of the CTTS population are different, supporting a different underlying emission mechanism between the two groups.

**Table 9.** Intrinsic X-ray fluxes and luminosity in the 0.35–7.5 keV band ($f_X$ and $L_X$) as derived from the EPIC pn spectral fits for the X-ray-bright PMS detected in L1551. Also given is the probability of constancy from the Kolmogorov-Smirnov (K-S) test for the XMM-*Newton* EPIC-pn light curves (shown in Fig. 3). The average flux level of the sources during the XMM-*Newton*, ROSAT and ASCA observations (when available) in the same bands as used by Carkner et al. (1996) is also reported. Carkner et al. provide ROSAT and ASCA fluxes over different energy ranges for each source; we report these energy ranges in the columns "$\Delta E_{\rm ROSAT}$" and "$\Delta E_{\rm ASCA}$" and provide the source flux over the same ranges derived from the fits to XMM-*Newton* spectral data. Energy ranges are in keV, flux units are $10^{-13}$ erg cm$^{-2}$ s$^{-1}$ and X-ray luminosity is in units of $10^{28}$ erg s$^{-1}$.

[a] average flux, the source underwent a large flare during ASCA observations.
[c] computed using the approximate relation of Carkner et al. (1996) 1 cts (ks)$^{-1} \simeq 3 \times 10^{28}$ ergs s$^{-1}$.
[d] unresolved from XZ Tau.
[e] as minimum flux level detectable by ROSAT we took the flux level of the weakest source detected by Carkner et al. (1996), MHO 4.
[f] in our data HL Tau was clearly visible in the image only for energies $\Delta E > 1.0$ keV (see text).

| Source | Name | $f_X$ | $L_X$ | K-S | $\Delta E_{\rm ROSAT}$ | $F_X$ XMM | $F_x$ ROSAT | $\Delta E_{\rm ASCA}$ | $F_X$ XMM | $F_X$ ASCA |
|---|---|---|---|---|---|---|---|---|---|---|
| S2 | MHO 5 | 0.66 | 15.6 | 0.79 | 0.2–2.0 | 0.34 | $\lesssim 0.4$ | 0.5–3.0 | 0.33 | – |
| S3 | V826 Tau | 23.0 | 534 | 0.13 10$^{-4}$ | 0.2–2.0 | 12.9 | 15.0 | 0.5–3.0 | 13.4 | 15.8[a] |
| S6 | V827 Tau | 4.0 | 93.8 | 0.37 10$^{-1}$ | 0.2–2.0 | 1.9 | 13.0 | 0.8–3.5 | 1.5 | 4.5 |
| S7 | V1075 Tau | 8.5 | 199 | 0.2 10$^{-25}$ | 0.1–2.0 | 4.1 | 5.3 | 0.5–2.5 | 4.1 | 4.9 |
| S14 | V710 Tau | 5.5 | 131 | 0.35 | 0.2–2.0 | 2.2 | 6.9[c] | 0.5–3.0 | 2.5 | – |
| S18 | XZ Tau | 55.3 | 1300 | $< 10^{-38}$ | 0.5–2.0 | 2.9 | $\sim 1.0$ | 0.5–2.0 | 2.9 | 15.0 |
| S19 | HL Tau | 10.0 | 235 | 0.74 10$^{-1}$ | 1.0–2.0[f] | 0.4 | [d] | 1.0–3.0 | 1.4 | [d] |
| S23 | MHO 4 | 0.31 | 7.34 | 0.54 | 0.2–2.0 | 0.2 | 0.4[c] | 0.5–3.0 | 0.2 | – |
| S25 | MHO 9 | 0.24 | 5.55 | 0.59 | 0.2–2.0 | 0.2 | $\lesssim 0.4$[e] | 0.5–3.0 | 0.2 | – |

The CTTS XZ Tau shows a very peculiar temporal variability, not reported previously. The inverse correlation between the X-ray flux and the absorbing column density points towards a "shadow" effect, compatible with a scenario in which the emission is shadowed by the accretion stream being brought between the emitting region and the observer, perhaps by the stellar rotation. The attendant implication is that the emission (to be effectively shadowed) must be spatially concentrated in a small region, suggestive of its being associated with the accretion spot(s), itself the most likely region on the stellar surface to be shadowed by the accretion stream. Kastner et al. (2002), based on the very peaked emission measure they derive for the high-resolution *Chandra* X-ray spectrum of the CTTS TW Hya also claim that its X-ray luminosity is likely to be (mostly) accretion driven.

While further observations are required to confirm the scenario described above, both the difference in metallicity and the difference in temporal variability point toward a difference in the underlying emission mechanism between the two populations. Such difference is in agreement with the result of Flaccomio et al. (2003), who find a systematic difference in the X-ray luminosity between CTTS and WTTS stars, with the former being less luminous at a given mass. The WTTS emit X-rays at a level close to (or at) the saturation level observed for older stars, supporting a scenario in which the underlying emission mechanism does not change as the star evolves from the WTTS stage onto the main sequence. The difference in the X-ray spectra and temporal behavior between CTTS and WTTS found here, together with the general lower emission levels reported by Flaccomio et al. (2003), are also indicative of a different underlying emission mechanism, which the shadowing observed for XZ Tau suggests to be (at least for a significant fraction) accretion driven.

*Acknowledgements.* GM, SS acknowledge the partial support of ASI and MIUR. This paper is based on observations obtained with XMM-*Newton*, an ESA science mission with instruments and contributions directly funded by ESA Member States and the USA (NASA). FF would like to thank I. Pillitteri for the support in the reduction of XMM-*Newton* data, J. J. Drake and the Harvard-Smithsonian Center for Astrophysics for the hospitality and for the help in analyzing the *Chandra* data, and L. Hartmann for the useful discussions.